\documentclass[a4paper]{article}
\usepackage{listings}
\usepackage{xcolor}
\usepackage{multirow}
\usepackage{jheppub}  
\usepackage[numbers]{natbib}
\usepackage{dcolumn}
\usepackage[english]{babel}
\usepackage[utf8x]{inputenc}
\usepackage[T1]{fontenc}
\usepackage{palatino}
\usepackage{indentfirst}
\usepackage{float}
\usepackage{amsmath}
\usepackage{afterpage}
\usepackage{graphicx, subcaption}
\usepackage[colorinlistoftodos]{todonotes}
\usepackage[colorlinks=true]{hyperref}
\usepackage{makeidx}
\newcommand{\be}{\begin{equation}}  
\newcommand{\ee}{\end{equation}}  
\newcommand{\bea}{\begin{eqnarray}}  
\newcommand{\eea}{\end{eqnarray}}  
\makeindex
\begin{document}

\vspace*{1.2cm}

\thispagestyle{empty}
\begin{center}
{\LARGE \bf Study of Collective Phenomena in Heavy-Ion Collisions
Using CMS Open Data}

\par\vspace*{7mm}\par

{

\bigskip

\large \bf Allan Eduardo Flores Godoi Ferreira$\mathbf{^1}$ \\
\large \bf César Augusto Bernardes$\mathbf{^2}$}

\bigskip

{\large \bf  E-Mail: allanfgodoi@gmail.com$\mathbf{^1}$}
{\large \bf  cesar.bernardes@unesp.br$\mathbf{^2}$}

\bigskip

{
Instituto de Física (IF) - Universidade Federal do Rio Grande do Sul (UFRGS) - Av. Bento Gonçalves, 9500, Porto Alegre - RS, Brazil$\mathbf{^1}$ \\
Núcleo de Computação Cientifíca (NCC) - Universidade Estadual Paulista "Júlio de Mesquita Filho" (UNESP) - Rua Dr. Bento Teobaldo Ferraz, 271, São Paulo – SP, Brazil$\mathbf{^2}$
}

\bigskip

{\it Presented at the Workshop of Advances in QCD at the LHC and the EIC, CBPF, Rio de Janeiro, Brazil, November 9-15 2025}


\vspace*{15mm}

\end{center}
\vspace*{1mm}

\begin{abstract}
aIn this work, we present preliminary results from a measurement of the recently proposed observable, $v_0(p_T)$, in lead-lead (PbPb) collisions at $\sqrt{s_{NN}}=2.76 \text{ TeV}$, using public data from the CMS Open Data portal. This observable is directly sensitive to radial flow and characterizes the transverse momentum ($p_T$) dependence of radial flow fluctuations, serving as probe of collective phenomena in the quark-gluon plasma (QGP) formed in heavy-ion collisions. Consistently with the ATLAS results at $\sqrt{s_{NN}}=5.02 \text{ TeV}$, we observe the three key features of collective radial flow: long-range correlations in pseudorapidity ($\eta$), a centrality-independent shape as a function of $p_T$, and factorization in $p_T$. The results presented in this work are generally compatible, within uncertainties, with the ATLAS measurements at $\sqrt{s_{NN}}=5.02\text{ TeV}$ reported in Phys. Rev. Lett. 136, 032301 (2026).
\end{abstract}

 
 \section{Introduction}

 In an ultrarelativistic heavy-ion collision, the final-state particle azimuthal angle ($\phi$) distribution can be described by a Fourier series \cite{hic-big}:
 
 \begin{equation}
     \frac{dN}{d\phi}=\frac{N}{2\pi}\left[ 1+2\sum^\infty_{n=1}v_n\cos[n(\phi-\Psi_{RP})] \right]
     \label{eq:fourier}.
 \end{equation}

 The harmonics $v_n$ carry information about the initial geometry of the collision and quantum fluctuations of the nucleons' positions inside the nuclei, and $\Psi_{RP}$ is the reaction plane angle. The independent term describes the radial flow, which is related to the spherically symmetric and collective increase of the medium velocity.

 As the medium evolves, pressure gradients in the initial state are transformed into a final momentum asymmetry that is measured by the detector. For this reason, it is useful to rewrite Eq. (\ref{eq:fourier}) in terms of harmonics that depend on the transverse momentum ($p_T$) \cite{hic2, hic-big, tcc-thierre, zhenyu-thesis, cesar-lapg, hanna-qm23}:

 \begin{equation}
     \frac{d^2N}{d\phi dp_T}=\frac{1}{2\pi}\frac{dN}{dp_T}\left[1+2\sum^\infty_{n=1}v_n(p_T)\cos[n(\phi-\Psi_{RP})] \right].
 \end{equation}


To build the newly proposed observable $v_0(p_T)$ \cite{chun-shen}, which describes the $p_T$ dependency of the radial flow fluctuations, we start from statistical definitions of covariance and correlation in an event-to-event context \cite{pheno1}.
Denoting by $[x]$ and $\langle x \rangle$, the mean of a quantity $x$ over a single event and over all events, respectively, and where $\delta [x] = [x] - \langle x \rangle$ denotes the fluctuation of this quantity, the covariance between $x$ and another quantity $y$ is given by:

\begin{equation}
    \text{cov}(x,y)= \langle (x - \langle x \rangle)(y - \langle y \rangle )\rangle = \langle \delta x \delta y \rangle.
\end{equation}

It is also possible to rewrite the covariance between these quantities as the product of their standard deviations and the Pearson correlation coefficient $\rho$, which measures the linear correlation between two quantities:

\begin{equation}
    \langle \delta x \delta y \rangle = \rho \sigma_x \sigma_y.
\end{equation}

Thus, the covariance between the normalized fractions of tracks in a given $p_T$ interval and the normalized mean $p_T$ of tracks in each event is expressed as \cite{ollitrault}:

\begin{equation}
    \frac{\langle \delta n(p_T) \delta [p_T] \rangle}{\langle n(p_T) \rangle \langle p_T \rangle} = \rho \sigma_n \sigma_{p_T}.
\end{equation}

The normalized $p_T$ standard deviation $\sigma_{p_T}$ is defined as $v_0 \equiv \sqrt{\langle(\delta [p_T])^2\rangle}/\langle p_T \rangle$ \cite{ollitrault, atlas-measurement}, a previously proposed and extensively investigated observable, while the remaining factor $\rho \sigma_n$ is the newly proposed observable $v_0(p_T)$, named in analogy to the azimuthal flow harmonics $v_n(p_T)$ \cite{ollitrault, atlas-measurement, atlas-qm25}. Therefore, this normalized covariance can be written as the product of two observables:

\begin{equation}
        \text{normalized covariance}\hspace{0.6cm} = 
        {\color{blue}\underbrace{v_0(p_T)}_{\text{$p_T$-dependent shape}}}
        \times
        {\color{red}\overbrace{v_0}^{\text{global amplitude}}}.
        \label{eq:factorization}
\end{equation}

However, the factorization described by Eq. (\ref{eq:factorization}) goes beyond the mathematical identity based on the Pearson correlation coefficient, because it assumes that $v_0$ is a common amplitude factor for all $p_T$ values, while the $v_0(p_T)$ observable governs the $p_T$ spectrum shape of the event. This also implies that fluctuations in the spectral shape are effectively decoupled from fluctuations in the overall magnitude of the radial flow. This supposition will be tested with data by performing the construction of the observables using different $p_T^{ref}$ ranges. Thus, in case of obtaining similar $v_0(p_T)$ values for different $p_T^{ref}$ ranges, the supposition will be validated \cite{atlas-qm25, atlas-is25}.

\begin{figure}[H]
        \centering
        \includegraphics[width=0.75\linewidth]{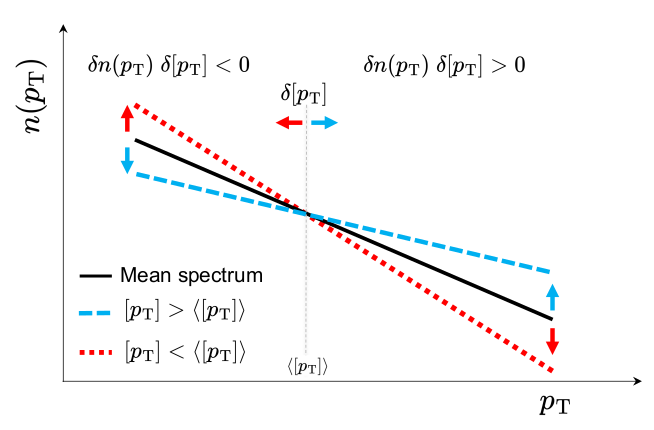}
        \caption{Scheme of the $p_T$ spectrum: the black curve represents the average spectrum over all events, while the blue and red curves represent events with a radial flow larger and smaller than the average, respectively. Events with a larger-than-average radial flow have a flatter spectrum, which is reflected in the covariance, that has a fixed zero-point at $p_T \approx \langle p_T \rangle$ \cite{atlas-measurement}.}
        \label{fig:spectrum}
\end{figure}

Physically, the event $p_T$ spectrum is modified by fluctuations in radial flow; therefore, it is possible to infer the radial flow of an event from the covariance of its $p_T$ spectrum using $v_0(p_T)$. Therefore, if an event has many high-$p_T$ tracks, where the radial flow is greater, $[p_T]$ tends to be greater than $\langle p_T \rangle$, resulting in a flatter $p_T$ spectrum. As in an event with few high-$p_T$ tracks, and consequently, lower radial flow, the spectrum becomes steeper. This 'seesaw' behavior of the spectrum around the average spectrum shows the existence of a pivot point in $p_T = \langle p_T \rangle$, which explains the characteristic zero-crossing observed in the $v_0(p_T)$ and can be observed in Fig. (\ref{fig:spectrum}) \cite{atlas-measurement}. 

In other words, we can investigate the radial flow looking at the spectrum shape, which is governed by $v_0(p_T)$. Taking the factorization shown in Eq. (\ref{eq:factorization}), the $v_0(p_T)$ observable can be obtained by dividing the measured normalized covariance, represented by the product $v_0(p_T)v_0$, by the global amplitude $v_0$.

\begin{equation}
    v_0(p_T) = \frac{v_0(p_T)v_0}{v_0}.
\end{equation}

A scaled version from the observable, $v_0(p_T)/v_0$, is also defined, it is independent on the size of the fluctuations and almost independent in the centrality of the collision, as will be shown in the results. It is obtained similarly as its original version:

\begin{equation}
    \frac{v_0(p_T)}{v_0} = \frac{v_0(p_T)v_0}{v_0^2}.
\end{equation}

Within the framework of a hydrodynamic description of the envolving system, this centrality independence can be understood as follows: while the magnitude of $v_0(p_T)$ and $v_0$ is governed by the system size at the initial stage of the collision for a fixed multiplicity, the shape, characterized by the ratio $v_0(p_T)/v_0$, depends solely in the hydrodynamic response of the medium, which is expected to be independent of centrality.

\section{Observables measurement}

This analysis is based on public data from lead-lead (PbPb) collisions at $\sqrt{s_{NN}}=2.76 \text{ TeV}$ collected with the CMS detector at LHC in 2011 \cite{ref-M}. A partial sample of approximately three million minimum-bias events is used, restricted to events with collision centrality greater than 50\%~\footnote{The CMS Collaboration has publicly released the PbPb 2011 data only for this centrality range; otherwise, additional centrality bins would have been included in the analysis.}. Collision centrality characterizes the degree of overlap of the two colliding lead nuclei \cite{ref-K}. A detailed description of the CMS detector can be found in Ref. \cite{ref-X}. Events and charged-particle tracks are selected according to standard CMS criteria \cite{ref-Y}. The selected tracks satisfy $0.5 < p_T <10 \text{ GeV}$ and $|\eta|<2.4$. 

The corrections for detector acceptance and tracking selection efficiency and fake rate are applied by weighting the tracks when we calculate $[p_T]$ and $n(p_T)$ as:

\begin{equation*}
    [p_T]=\frac{\sum_iw_ip_{T_i}}{\sum_iw_i},
\end{equation*}

and

\begin{equation*}
    n(p_T)=\frac{\sum_i^{p_T}w_i}{\sum_iw_i}.
\end{equation*}

Where $\Sigma_i^{p_T}$ denotes the sum over all tracks in a small $p_T$ bin. The weights are derived using Monte Carlo simulations by comparing charged particles in the generator level with reconstructed tracks \cite{ref-K, ref-Y}.

To suppress non-flow correlations (e.g. jets and resonances), we adopt a 2-subevent method, where the dataset is split into two symmetric subevents separated by a zero-centered $\eta_{gap}$. The subevents A and B are defined within the intervals $-2.4 \leq \eta_A \leq -\eta_{gap}/2$ and $\eta_{gap}/2 \leq \eta_B \leq 2.4$ \cite{subevents, ollitrault, ollitrault-slides, atlas-measurement}.

Following the 2-subevent method, the variance of $[p_T]$ can be expressed as the product of its variances from each subevent and the covariance $\langle\delta n(p_T)\delta[p_T]\rangle$ as the cross-mean between the subevents. Therefore, we can experimentally obtain $v_0$ and $v_0(p_T)v_0$ as:

\begin{equation}
    v_0= \frac{\sqrt{\langle (\delta[p_T])^2 \rangle}}{\langle p_T \rangle} = \frac{\sqrt{\langle \delta[p_T]_A\delta[p_T]_B\rangle}}{\langle p_T \rangle},
\end{equation}

and

\begin{equation}
    v_0(p_T)v_0=\frac{\langle\delta n(p_T)\delta[p_T]\rangle}{\langle n(p_T)\rangle\langle p_T\rangle}=\frac{1}{2}\frac{\langle \delta n_A(p_T)\delta[p_T]_B+\delta n_B(p_T)\delta[p_T]_A\rangle}{\langle n(p_T)\rangle\langle p_T\rangle}.
\end{equation}

To verify the factorization assumption, we test different $p_T^{ref}$ ranges in the construction of the observables, expecting similar results for each choice \cite{atlas-measurement, atlas-qm25, atlas-is25}. The $p_T^{ref}$ range is employed solely for the evaluation of $\delta[p_T]$ and $\langle p_T \rangle$; conversely, $\delta n(p_T)$ and $\langle n(p_T)\rangle$ are derived for each analysis bin within the range $0.5 < p_T < 10 \text{ GeV}$.

\section{Results}

The following preliminary results were obtained using approximately three million minimum-bias events. We also compare them with the ATLAS measurement using PbPb collisions at $\sqrt{s_{NN}}=5.02 \text{ TeV}$ to investigate how the energy difference affects the observables. The error bars in the plots represent the statistical uncertainties of the measurements, which were calculated using a Poisson Bootstrap method \cite{bootstrap}.

\begin{figure}[H]
    \centering
    \includegraphics[height=6.0cm]{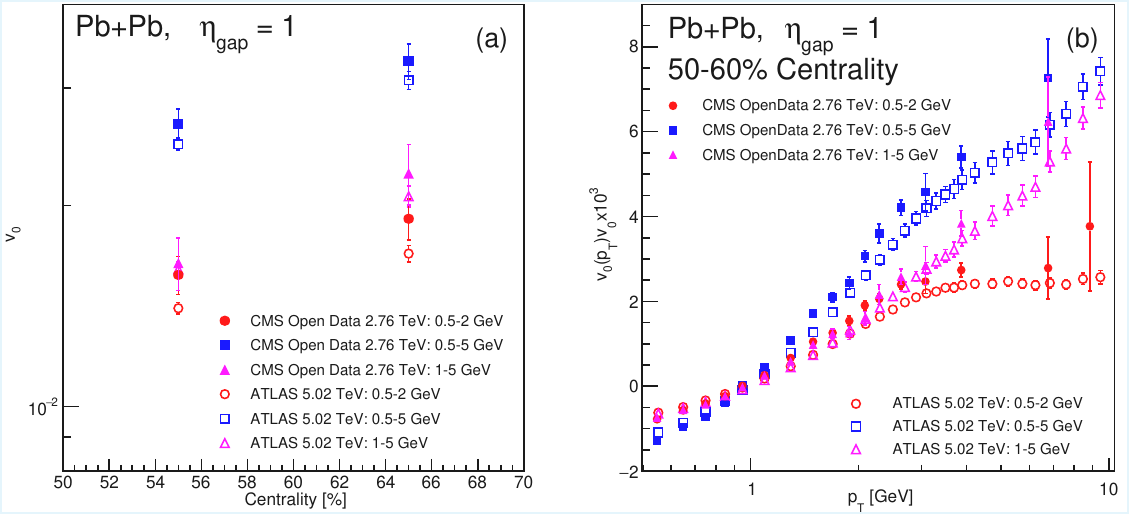}
    \caption{$v_0$ dependence on centrality \textbf{(a)} and the normalized covariance $v_0(p_T)v_0$ \textbf{(b)}, both measured in different $p_T^{ref}$ ranges.}
    \label{fig:v0-v0ptv0}
\end{figure}

In Fig. (\ref{fig:v0-v0ptv0}a) we can observe how $v_0$ depends on centrality and on $p_T^{ref}$, note that in this analysis we will be covering only two centrality ranges, since this is the available data. While in Fig. (\ref{fig:v0-v0ptv0}b) we see the shape of the normalized covariance at different $p_T^{ref}$ for 50-60\% centrality, the 'seesaw' behavior can be observed while all $p_T^{ref}$ measurements have a common zero-crossing point. Due to the $\langle p_T \rangle$ for the two collision energies are similar, they cross the zero almost at the same $p_T$ value. The magnitudes for the $v_0$ and $v_0(p_T)v_0$ seem to be systematically above for the $2.76 \text{ TeV}$ compared to $5.02 \text{ TeV}$, but in most of the bins they are compatible within uncertainties.

\begin{figure}[H]
    \centering
    \includegraphics[height=6.0cm]{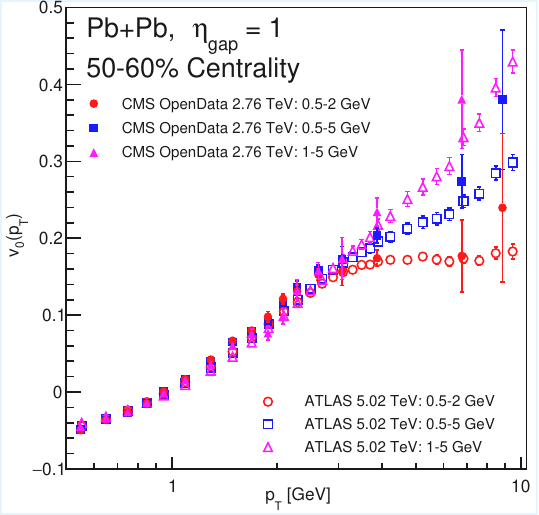}
    \caption{$v_0(p_T)$ observable in different $p_T^{ref}$ ranges and $\eta_{gap}=1$ in 50-60\% centrality.}
    \label{fig:v0pt-refs}
\end{figure}

By scaling the normalized covariance, we obtain $v_0(p_T)$ in different $p_T^{ref}$ ranges for 50-60\% centrality. As we can observe in Fig. (\ref{fig:v0pt-refs}), the observable values do not depend on $p_T^{ref}$ in the low-$p_T$ hydrodynamical regime, validating the factorization assumption and also indicating a collective behavior of the medium. In addition, the results got closer to ATLAS values at $5.02 \text{ TeV}$, indicating a possible cancellation of initial stages effects in the two collision energies.

\begin{figure}[H]
    \centering
    \includegraphics[height=6.0cm]{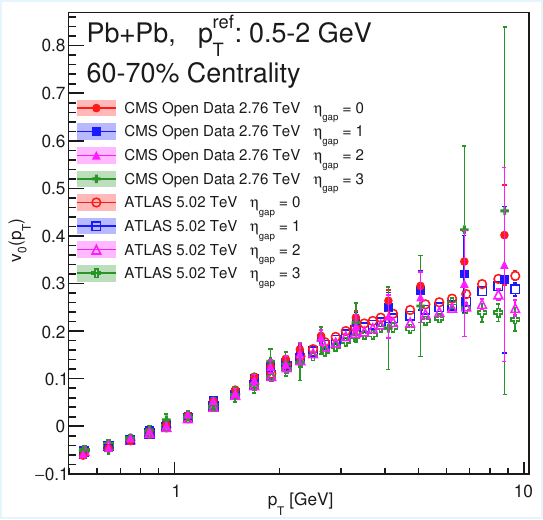}
    \caption{$v_0(p_T)$ in different $\eta_{gap}$ values and $p_T^{ref} \in [0.5,2] \text{ GeV}$ in 50-60\% centrality.}
    \label{fig:v0pt-etas}
\end{figure}

We also compute $v_0(p_T)$ with different $\eta_{gap}$ values in Fig. (\ref{fig:v0pt-etas}), where a weak dependence on $\eta_{gap}$ is observed, specially in low-$p_T$ regime, meaning that the observable is not sensitive on the detector region from the measurement. This behavior in different $\eta$ regions characterizes a long-range correlation, which also indicates collectivity.

\begin{figure}[H]
    \centering
    \includegraphics[height=6.0cm]{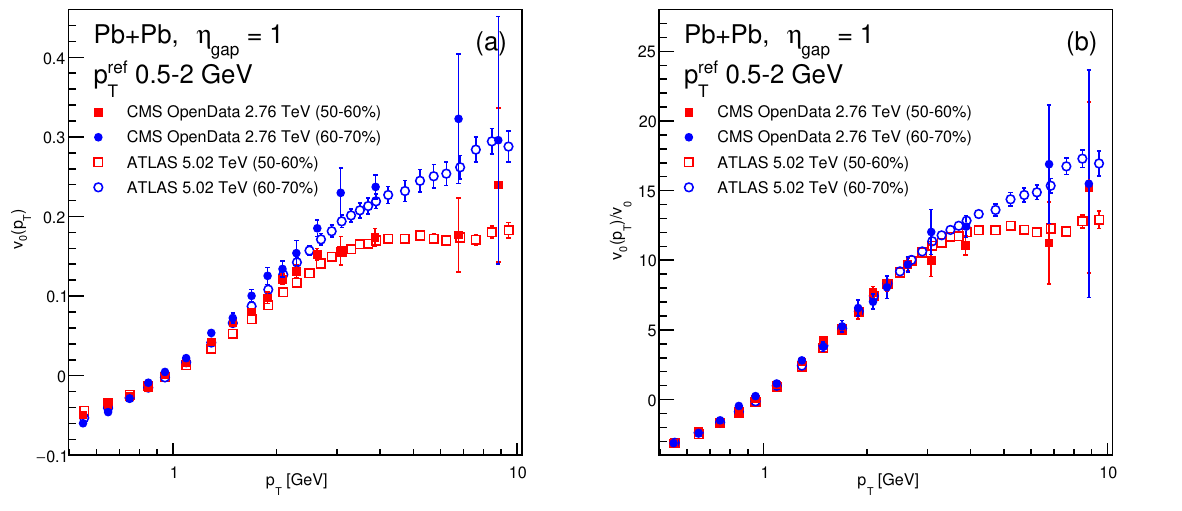}
    \caption{$v_0(p_T)$ and $v_0(p_T)/v_0$ at default parameters in different centrality ranges.}
    \label{fig:plot-main}
\end{figure}

Setting $\eta_{gap}=1$ and $p_T^{ref}$ range within $0.5\text{-}2.0~\mathrm{GeV}$ as default parameters, in Fig. (\ref{fig:plot-main}) we observe a similar trend to our measurement compared with that obtained by ATLAS in our two available centrality ranges. Also, the centrality independence of $v_0(p_T)/v_0$ can be observed at low-$p_T$.

\section{Summary}

In this proceedings, we present a measurement of the observable $v_0(p_T)$, which is sensitive to radial flow fluctuations as function of transverse momentum ($p_T$). The measurement is performed in lead-lead (PbPb) collisions at $\sqrt{s_{NN}}=2.76 \text{ TeV}$, using publicly available data collected by the CMS experiment at the LHC in 2011. The results are compared with previous ATLAS measurements in PbPb collisions at $\sqrt{s_{NN}}=5.02 \text{ TeV}$ and are found to be compatible within uncertainties. As a next step, we plan to incorporate systematic uncertainties and process additional data to increase the event sample, thereby reducing statistical uncertainties. Furthermore, we aim to compare our experimental results with theoretical predictions at $\sqrt{s_{NN}}=2.76 \text{ TeV}$.
 
\section*{Acknowledgements}

We thank the workshop organizing comittee, the Heavy-ion Group from São Paulo Research and Analysis Center (SPRACE), the Experimental Group of the CMS Collaboration (GECMS-UFRGS), the Particle Analysis and Simulation Group (GASP-UFRGS) and the Conselho Nacional de Desenvolvimento Científico e Tecnológico (CNPq) for supporting this work with a Scientific Initiation fellowship under Grant No. 159870/2025-9.

\bibliographystyle{unsrt}
\bibliography{bib}

\end{document}